\documentclass[12pt,a4paper]{article}

\usepackage{amsmath}

\begin{document}

\bibliographystyle{plain}

\begin{titlepage}

\hfill hep-th/0307164 

\hfill LTH 584

\vspace{1.5in}

\begin{center}

{\Large{Response to `Comments on the U(2) ADHM two-instanton'}} \\
\bigskip
\bigskip
\bigskip

{\large{N.\ B.\ Pomeroy}} \\

\bigskip

{\emph{Theoretical Physics Division, Department of Mathematical Sciences, \\ University of Liverpool, Liverpool, L69 3BX, U.K.}} \\

\bigskip

\texttt{nbp@amtp.liv.ac.uk}

\end{center}

\vspace{0.75in}

\begin{center}
{\Large{Abstract}} 
\end{center}

We respond to `Comments on the $U(2)$ ADHM two-instanton' [Y.~Tian, Phys.\ Lett.\ B {\bf 566} (2003) 183]. 

\end{titlepage}

The author of \cite{tian} claims that for $N=2$ the $U(N)$ two-instanton gauge field configuration given in \cite{pomeroy} by explicitly resolving the ADHM instanton constraints \cite{adhm,cftg} is incorrect. In response, we highlight some of the shortcomings of the $U(2)$ ADHM two-instanton solution presented in \cite{tian}, and re-examine the $U(2)$ ADHM two-instanton solution which we presented in \cite{pomeroy} in view of the comments upon it given in \cite{tian}. \\
 
Firstly, the $U(2)$ two-instanton solution presented in \cite{tian} does not take into account the fact that the full auxiliary symmetry of the $U(2)$ ADHM construction with topological charge $k=2$ is $U(2)$ and not $SU(2)$. The $SU(2)$ factor within the global $U(2) \simeq U(1) \times SU(2)$ transformations which must be performed to fix the auxiliary symmetry is accounted for, but the $U(1)$ factor is not. The $U(1)$ factor within the $U(2)$ global gauge transformations will act trivially upon the submatrix $a^{\prime}$ consisting of the submatrices $B_{1}$ and $B_{2}$, in the notation of \cite{tian}. However, the $U(1)$ factor may act upon the submatrices $I$ and $J$ of the $U(2)$ ADHM two-instanton solution presented in \cite{tian} non-trivially. This is not addressed, nor accounted for, for the $U(2)$ ADHM two-instanton solution presented in \cite{tian}, as only the action of the $SU(2)$ transformation upon the submatrices $I$ and $J$ is  considered, as in Eq.~(6) of \cite{tian}. The aforementioned global $U(1)$ gauge transformations which should act upon the submatrices of the $U(2)$ ADHM two-instanton solution presented in \cite{tian} may add additional parameters to the that solution, thereby making the $U(2)$ ADHM two-instanton solution presented in \cite{tian} a special, rather than general, solution. An explicit form for the global gauge transformation used to fix the $U(2)$ auxiliary symmetry of the $U(2)$ ADHM two-instanton presented in \cite{pomeroy} was given in \cite{pomeroy}, which included both the action of the $SU(2)$ factor and of the $U(1)$ factor within the $U(2)$ group upon all of the ADHM submatrices of the ADHM matrix $a$. \\

Secondly, we consider the contradiction alleged by the author of \cite{tian} to be present in the $U(2)$ ADHM two-instanton solution presented in \cite{pomeroy}. Equations (35) and (46) of \cite{pomeroy} are not necessarily in contradiction. Rather, given the form of the $U(2)$ ADHM two-instanton submatrices in Eqs.~(47,48) of \cite{pomeroy}, where the $U(2)$ auxiliary symmetry has been fixed, if one demands consistency between these equations and the form of the fixed $U(2)$ auxiliary symmetry as given in the Appendix of \cite{pomeroy}, so that $U_{1}<0$ (we note that $U_{1}>0$ is not permitted, in order that the $U(2)$ two-instanton solution with $U(2)$ auxiliary symmery fixed as in \cite{pomeroy} is a valid solution), one derives the following condition upon the parameters contained within the $U(2)$-auxiliary symmetry-fixed $U(2)$ ADHM two-instanton solution in \cite{pomeroy}:
\begin{equation}
\frac{2|x_{0}|^{2}}{(|x_{0}|^{2}+|x_{1}|^{2})|x_{1}|^{2}}[ \bar{x}_{0}x_{1}U_{2}U_{z}+x_{0}\bar{x}_{1}\bar{U}_{2}\bar{U}_{z}] \ > \ 0.
\label{eq:bound}
\end{equation}
The inequality in Eq.~(\ref{eq:bound}) is specific to the $U(2)$ ADHM two-instanton with the $U(2)$ symmetry fixed as in \cite{pomeroy}, and does not constitute a generic defect of the $N=2$ case of the $U(N)$ ADHM two-instanton solution given in \cite{pomeroy}, as there exist choices of transformations other than those in the Appendix of \cite{pomeroy} with which to fix the $U(2)$ auxiliary symmetry when $N=2$. For the specific choice of transformations used to fix the $U(2)$ auxiliary symmetry given in the Appendix of \cite{pomeroy}, we admit that the inequality in Eq.~(1) does represent an additional constraint upon the parameters of the $U(2)$ ADHM two-instanton solution, so that the parameters in this particular form of the solution are not completely independent. However, Eq.~(\ref{eq:bound}) does not reduce the number of parameters in the solution as it is an inequality. The inequality in Eq.~(\ref{eq:bound}) serves as a bound on the parameters in the solution, but the $U(2)$ ADHM two-instanton solution in \cite{pomeroy} is still parameterized by sixteen real parameters. In physical applications, this bound may naturally be satisfied. We maintain that other choices of transformations which fix the auxiliary $U(2)$ symmetry of the $N=2$ solution exist which do not create the bound in Eq.~(\ref{eq:bound}). Hence there is no contradiction, only a bound imposed upon the parameters of the solution. \\

Thirdly, the quantity $|x|^{2}$ defined in Eq.~(33) of \cite{pomeroy} was not specified by us in \cite{pomeroy} as the separation between the two constituent one-instantons of the $U(N)$ ADHM two-instanton solution in \cite{pomeroy}. The quantity $|x|^{2}$ was defined only for convenience, and we note that in the original paper \cite{pomeroy}, a typographical error was made. In agreement with the author of \cite{tian}, the quantity $|x|^{2}$ should be defined as: 
\begin{equation}
|x|^{2} \ \equiv \ |x_{0}|^{2}+|x_{1}|^{2},
\end{equation}
for which the solution presented in \cite{pomeroy} does indeed satisfy the ADHM constraints for gauge group $U(N)$ and $k=2$. Equation (2) above corrects Eq.~(33) of \cite{pomeroy}. The quantity $|x|$ is the separation to be expected between two points located at $(-\tfrac{1}{2}x_{0},-\tfrac{1}{2}x_{1})$ and $(\tfrac{1}{2}x_{0},\tfrac{1}{2}x_{1})$ on a flat Euclidean complex space of complex dimension two, in accordance with the form of the submatrices specifying the general $U(N)$ ADHM two-instanton given in Eqs.~(38,39) of \cite{pomeroy}. Therefore the quantity $|x|^{2}$ is indeed associated with the separation between the two constituent $U(N)$ ADHM one-instanton solutions within the $U(N)$ ADHM two-instanton configuration given in \cite{pomeroy}. We thank the author of \cite{tian} for bringing this typographical error to our attention. \\

Fourthly, we examine the dilute instanton gas limit, also known as the completely clustered limit, of the $U(2)$ ADHM two-instanton solution presented in \cite{tian}. In Eq.~(14) of \cite{tian}, the author of \cite{tian} obtains the dilute instanton gas limit by taking the separation between the constituent $U(2)$ ADHM one-instantons to infinity. We assume that the ADHM submatrix $a$ in Eq.~(14) of \cite{tian} represents the ADHM data of a general $U(2)$ ADHM two-instanton solution of arbitrary $U(2)$ gauge iso-orientation. However, one of the $U(2)$ ADHM one-instanton submatrices within the decomposition in Eq.~(14) does not possess an arbitrary $U(2)$ gauge iso-orientation: in the notation of \cite{tian}, the parameters $\{ l,m,\theta \}$, which presumably should also be present to specify the $SU(2)$ gauge iso-orientation of the $U(2)$ ADHM one-instantons, assume special values for one of the $U(2)$ ADHM one-instantons. In contrast, the other $U(2)$ ADHM one-instanton submatrix possesses arbitrary $SU(2)$ gauge iso-orientation parameters $\{ l,m,\theta \}$. Furthermore, this implies that one of the one-instanton submatrices contains fewer parameters than the other one-instanton submatrix, which should not be the case. (The total number of real parameters for the $U(2)$ auxiliary-symmetry-fixed $U(2)$ ADHM two-instanton is sixteen. Within the dilute instanton gas limit, the two constituent $U(2)$ ADHM one-instanton solutions which arise should possess eight parameters each.) The $U(2)$ ADHM two-instanton solution presented in \cite{pomeroy} does not exhibit this deficiency in the number of parameters when the dilute instanton gas limit is taken. Moreover, the $U(2)$ ADHM two-instanton solution in \cite{pomeroy} possesses an arbitrary $U(2)$ gauge iso-orientation, which remains arbitrary, though possibly correlated, for each of the two $U(2)$ ADHM one-instanton solutions in the dilute instanton gas limit. In this way, both of the constituent $U(2)$ ADHM one-instantons obtained in the dilute instanton gas limit of the $U(2)$ ADHM two-instanton solution in \cite{pomeroy} each possess their own $U(2)$ gauge iso-orientations and the correct number of real parameters for general $U(2)$ ADHM one-instantons. \\

The author of \cite{tian} also considers the 't Hooft limit for the $U(2)$ ADHM two-instanton solution presented in \cite{tian}, in which the constituent $U(2)$ ADHM one-instantons possess the same gauge iso-orientation. It is not clear how one can take the 't Hooft limit for the $U(2)$ ADHM two-instanton solution presented in \cite{pomeroy}. This is because the physical interpretation of the parameters within the solution is not sufficiently complete to enable one to specify that the two constituent one-instantons in the solution possess the same gauge iso-orientation. However, we believe that this does not indicate that the $U(N)$ ADHM two-instanton solution given in \cite{pomeroy} is incorrect. The complexity of the $U(N)$ ADHM two-instanton solution given in \cite{pomeroy} does not yet permit a completely transparent identification of the gauge iso-orientation parameters within in it, though, as explained in \cite{pomeroy}, they are contained within the ADHM submatrices $u_{1}$ and $u_{2}$ used in the $U(N)$ ADHM two-instanton solution presented in \cite{pomeroy}. The general formalism for $U(N)$ ADHM multi-instantons developed in \cite{dhkm} is expected to permit such an identification in future. We therefore cannot yet comment on the 't Hooft limit of the $U(2)$ ADHM two-instanton solution presented in \cite{pomeroy}. Currently, there is no evidence that the 't Hooft limit does not exist for the $U(2)$ ADHM two-instanton solution presented in \cite{pomeroy}. \\

We maintain that the $U(N)$ ADHM two-instanton solution given in \cite{pomeroy} is correct. Any discrepancies for the case $N=2$, as claimed by the author of \cite{tian}, are due to the choice made in \cite{pomeroy} for the fixing of the $U(2)$ auxiliary symmetry, which is required to render the solution physical. Since these choices are arbitrary and not unique, we presume that other global $U(2)$ gauge transformations may be performed so that any such discrepancies can be removed. However, finding such transformations in order to do this is a difficult task, which we intend to address elsewhere. \\

{\Large{\textbf{Acknowledgements}}} \\

We wish to acknowledge helpful discussions with Prof.\ D.\ R.\ T.\ Jones and Dr.\ I.\ Jack.

\end{document}